\documentclass[sigconf]{acmart}
\AtBeginDocument{%
  \providecommand\BibTeX{{%
    \normalfont B\kern-0.5em{\scshape i\kern-0.25em b}\kern-0.8em\TeX}}}

\setcopyright{acmcopyright}
\copyrightyear{2018}
\acmYear{2018}
\acmDOI{XXXXXXX.XXXXXXX}

\acmConference[Conference acronym 'XX]{Make sure to enter the correct
  conference title from your rights confirmation emai}{June 03--05,
  2018}{Woodstock, NY}
\acmPrice{15.00}
\acmISBN{978-1-4503-XXXX-X/18/06}

\begin{document}

\title{Social Dynamics of Online Pro-Eating Disorder  Communities}

\author{Kristina Lerman$^{*,1, 2}$, Aryan Karnati$^2$, Shuchan Zhou$^2$, Siyi Chen$^2$, Sudesh Kumar$^2$, Zihao He$^2$, Joanna Yau$^2$, Abigail Horn$^{1,2}$}
\email{lerman@isi.edu}
\affiliation{%
\institution{1. USC Information Sciences Institute\\
2. University of Southern California}
\country{USA}
}

\renewcommand{\shortauthors}{Lerman, et al.}

\begin{abstract}
The rise in eating disorders, a mental health condition with serious medical complications, has been linked to the proliferation of idealized body images on social media. However, the link between social media and eating disorders is more complex, with online platforms potentially enabling harmful behaviors by linking people to ``pro-ana'' communities that promote eating disorders.
We conceptualize the growth of harmful pro-ana  communities as a process of online radicalization. We show that a model of radicalization explains how individuals are driven to conversations about extreme behaviors, like fasting, to achieve the ``thin body'' goal, and how these conversations are validated by pro-ana communities.
By facilitating social connections to like-minded others, a shared group identity and emotional support, social media platforms can trap individuals within toxic echo chambers that normalize disordered eating behaviors and other forms of self-harm. We also show that while harmful messages are easily accessible online, they are also easy to distinguish from the non-harmful messages, thereby enabling content moderation.
Characterizing and quantifying the role of online communities in amplifying harmful conversations will support the development of strategies to mitigate their impact and promote better mental
health.

\end{abstract}

\ccsdesc[500]{Human-centered computing}
\ccsdesc[300]{Information systems}
\ccsdesc[500]{Applied computing~Psychology}
\ccsdesc[500]{Human-centered computing~Empirical studies in collaborative and social computing}

\keywords{twitter, eating disorders, mental health, radicalization}



\maketitle

\section{Introduction}
\textcolor{red}{[\textbf{Warning: This paper discusses eating disorders and self-harm, which some  may find distressing.}]}

Eating disorders represent a serious mental health condition characterized by obsessive thoughts and unhealthy behaviors around food, eating, body weight and shape. The condition, which includes anorexia, bulimia, and binge eating disorder, affects 24 million people in the US. Many sufferers have their health, daily routines and quality of life severely compromised due to medical complications, which also impose a high toll on family members and other caregivers~\cite{vanHoeken2020review}.  Eating disorders are also among the deadliest of all mental health conditions,  with over 10,000 deaths each year.
To the alarm of public health experts and mental health professionals, the prevalence of eating disorders has more than doubled in recent years, growing from 3.5\% in the 2000-2006 period to 7.8\% in the 2013-2018 period~\cite{Galmiche2019}. 
The Covid-19 pandemic has further exacerbated the condition~\cite{nutley2021impact}, particularly among girls and young women~\cite{Hartman2022}. According to the CDC, the number of weekly visits to emergency rooms for adolescent girls struggling with eating disorders has doubled in 2021 compared to 2019~\cite{radhakrishnan2022pediatric}. These trends, along with the growing gender disparity, call for urgent solutions.

Researchers have linked the rise in eating disorders to the proliferation of idealized body images on visual social media platforms like Instagram and TikTok that promote the ``thin ideal'' and the ``muscular ideal''~\cite{fardouly2016social,marks2020pursuit}. Exposure to idealized images can be distressing, inviting negative comparisons and pressure to conform to unrealistic beauty standards. This, in turn, makes individuals feel worse about themselves and contributes to  negative  body image, a known risk factor for developing eating disorders ~\cite{choukas2022perfectstorm}.
However, the link between social media and eating disorders goes beyond negative body image. The risks are two-fold. First, social media search and recommendation algorithms may provide a gateway for youth searching for health, nutrition and fitness information to discover harmful content that promotes extreme dieting and over-excercizing. Second, individuals may stumble upon ``pro-ana'' communities that glorify thinness and view anorexia as an aesthetic rather than a serious medical condition~\cite{Ging2018}. While such communities may provide validation, a safe space to vent, and emotional support  to individuals who often feel stigmatized~\cite{oksanen2016proanorexia,YeshuaKatz2013stigma}, they may ultimately harm them by trapping them within worldviews that normalize disordered behaviors and delay recovery~\cite{pater2016hunger,pater2017defining}.
Although such pro-ana communities have existed since the early days of the web in the form of blogs and message forums~\cite{Ging2018}, social media has vastly expanded their reach and accessibility~\cite{choukas2022perfectstorm}. According to one recent study, discussions about  eating disorders and other forms of self-harm on Twitter have quintupled in less than a year~\cite{Goldenberg2022ncri}. \textit{By helping users find one another within pro-ana spaces, social media has supercharged their growth.} 

Our paper attempts to explain these trends by casting the growth of online communities that promote eating disorders---\emph{pro-ED communities}---as a radicalization process. This framing allows us to draw on rich literature and theory that explain how {online group dynamic processes} can radicalize individuals by linking them to like-minded others who push them towards more extreme beliefs and behaviors.
Prior works have shown similar group dynamics amplify political polarization~\cite{barbera2020social}, hate speech~\cite{schmitz2022quantifying}, misinformation~\cite{chou2018addressing} and  conspiracy movements~\cite{wang2022identifying}. However, the link between online radicalization and mental health disorders has not be widely recognized.

Our approach is inspired by the \textit{3N model of radicalization}~\cite{Kruglanski2022}. Originally developed to explain how terrorists and violent extremists are formed~\cite{belanger2019radicalization,Kruglanski2014}, this theoretical framework has been applied to non-violent behaviors such as joining the QAnon conspiracy~\cite{wang2022identifying}. The model explains how loss of personal significance (\textit{Need} for significance) moves individuals to justify   extreme behaviors (create \textit{Narratives}) that enable them to achieve their goals and thereby regain significance. These narratives are validated by communities (\textit{Networks}) that connect individuals to like-minded others, while isolating them from opposing worldviews.
We operationalize key concepts of the 3N model and link them to data using a large corpus of online eating disorders-related conversations. Although alternative theoretical lenses have been applied to pro-ED online spaces~\cite{pater2017defining,choukas2022perfectstorm}, they fail to account for group dynamics. 







Our study focuses on Twitter, which has an  active---and quickly growing~\cite{Goldenberg2022ncri}---eating disorders community. 
Twitter does not forcefully moderate harmful pro-ED content~\cite{Sukunesan2021}.  Unlike TikTok and Instagram, which returned links to National Eating Disorder Association resources when searching for ED-related topics, Twitter returned harmful posts at the time that data was collected. We show that content related to eating disorders is deeply integrated within mainstream topics such as ``diet'' and ``weightloss.'' Moreover, due to the intersections between various mental health disorders, individuals viewing pro-ED messages are often exposed to self-harm behaviors like cutting. Thus, eating disorders may be a gateway to other  behaviors and beliefs that erode mental health.

We empirically ground the 3N model by operationalizing its constructs and quantifying them using Twitter data. For example, the model predicts that individuals may regain personal significance by adopting a group identity. We show that a large share of accounts  posting about eating disorders adopt a  collective identity through the use of self-defining hashtags~\cite{barron2022quantifying}, such as ``edtwt'' (eating disorders Twitter) or ``shtwt'' (self-harm Twitter). While self-defining hashtags were studied in the context of social justice movements~\cite{chang2022justiceforgeorgefloyd}, their importance in mental health conditions has not been as widely acknowledged.
Next, we analyze pro-ED conversations and find that, consistent with the 3N model, many promote and justify disordered eating behaviors. Specifically, they motivate extreme behaviors through ``inspirational'' images and accountability posts, provide tips to manage ED symptoms or hide them from caregivers, as also found by other researchers~\cite{chancellor2016post,pater2016hunger,chancellor2016quantifying}.
Finally, we characterize the quality of interactions between members of pro-ED spaces to demonstrate that pro-ED communities provide social and emotional support. Using state-of-the-art language models, we measure emotions expressed in  pro-ED tweets to show that positive emotions like joy dominate. To best of our knowledge this is the first large scale analysis of emotions of ED-related conversations.

The online radicalization framework not only allows us to better understand the social dynamics through which pro-ED spaces grow, but also suggests strategies to limit their growth. These range from content moderation to limit exposure to harmful content to banning problematic communities and users. Focusing on content moderation, we show that it is feasible to automatically distinguish harmful tweets from the more mainstream content.

\textbf{Contributions of this paper: }
Our contributions are three-fold. First, we audit a sample of Twitter conversations about nutrition and fitness and show that \textit{content promoting eating disorders is widely accessible} via popular topics like diet and weightloss.
Second, we use the framework of online radicalization to describe the social dynamics of pro-ED communities. This conceptualization allows us to cast phenomena that were studied separately in the past --- eating disorders as an identity, the sharing of pro-ED tips and motivating ``thinspo'' imagery, providing emotional support --- as part of \emph{the same underlying process}.
Also, by uniting such seemingly disparate processes as political polarization, online extremism,  conspiracy theories and mental health under the umbrella of radicalization hints at universal human needs that drive them.
As our third contribution we demonstrate that it is possible to automatically distinguish harmful tweets from other conversations about diet and fitness.


Our findings highlight the dangerous ``harm spiral'' created by the technological affordances of social media: \emph{harmful content is easy to discover and group dynamics encourage people to stay engaged, which exposes them to more harmful content}.
Note that processes similar to radicalization can create ``virtuous spirals'' that result in social justice movements~\cite{chang2022justiceforgeorgefloyd} or online support groups for marginalized identities. However, the fundamental asymmetry of human cognition that makes us pay more attention to negative information~\cite{baumeister2001bad} make the harm spirals more pervasive and dangerous. However, since radicalization is a feedback loop, even small changes can have large downstream effects. Specifically, content moderation, which would automatically limit exposure to harmful tweets by making them harder to find, could have a big impact on slowing the growth of pro-ED communities.

\textbf{Ethics:}
This research touches on highly sensitive topics related to mental health, which calls for extra precautions to minimize risks to study subjects as well as researchers.
All data used for this study is public and collected following Twitter's terms of service. 
To minimize privacy risks, identifiable information was removed and analysis was carried out on aggregated data. As a result, we believe that the risks of negative outcomes due to the use of these data are trivial.
To minimize risks to researchers, we did not collect images and regularly met with the research team to identify potential sources of distress.


One potential objection is that using the term ``radicalization'' to describe people with eating disorders is wrong, since they are not violent, and moreover, risks stigmatizing an already vulnerable population. As a counterargument, we offer that the radicalization paradigm has been used to describe both violent (terrorism) and non-violent (conspiracies) phenomena. Moreover, uniting these disparate phenomena under the umbrella of radicalization is evidence for the universality of online social dynamics.
Another potential objection is whether pro-ED speech should be moderated at all. We argue that pro-ED speech normalizes harmful behaviors that at minimum delay recovery but potentially cause more distress to people struggling with ED, their families and caregivers. Disrupting the group dynamics of polarization by moderating pro-ED speech would create ripple effects that reduce pro-ana groups.

It is important to note that our study cannot discriminate between individuals with eating disorders and those who are merely discussing disordered behaviors and beliefs. Nor do we have evidence that joining pro-ana conversations results in people developing eating disorders or exacerbating an existing illness. We argue, however, that such communities are harmful in that they normalize and promote ED behaviors, potentially delaying recovery for individuals struggling with ED.
With these risks in mind, our study identifies a new concern in the mental health crisis, specifically, online communities which trap individuals within toxic echo chambers that entrench and amplify extreme views. Better understanding the role of social media platforms in promoting eating disorders content and communities will inform the development of better strategies to enhance mental health.

\section{Related Works}



Eating disorders have complex biopsychosocial etiology, with contributing biological factors, such as genetics and infections~\cite{aman2022prevalence}, as well as psychological comorbidities, such as anxiety and perfectionism.
In addition, social mechanisms such as peer effects~\cite{allison2014anorexia} and exposure to idealized body images in the media, contribute to eating disorders. For example,   before the introduction of Western TV programming on Fiji in 1995, purging as a means of controlling weight was virtually unknown, but quickly grew afterwards~\cite{becker2004television}.

\subsection{Social Media and Body Image Concerns}
One of the best studied area is the relationship between social media and body image concerns, a key risk factor for developing depressive symptoms and eating disorders ~\cite{choukas2022perfectstorm}. Body image concerns arise when a person feels anxious, distressed, or self-conscious about their body weight, shape or appearance.
Studies have shown that social media use contributes to body image concerns in several ways. First, content on social media sites, especially image-based platforms like Instagram, Snapchat and Tiktok, features images of bodies, often heavily edited, that promote the `thin ideal'. 
Studies have shown that when users are exposed to this type of content, they  compare themselves to the idealized body images and as a result, feel worse about their own body and appearance~\cite{Saiphoo2019, fardouly2016social,choukas2022perfectstorm}.
Similar to traditional media, exposure to the ``thin ideal'' by influencers on social media can provoke upward social comparisons, where individuals compare themselves to people whom they perceive to be better than themselves ~\cite{harriger2022rabbithole}. In contrast to television and magazines, social media also allows adolescents to compare themselves to peers, which provides more salient and potent point of negative social comparison~\cite{Festinger}. What's more, adolescents are often not aware that images are frequently heavily edited~\cite{marks2020pursuit}, and as a result they normalize distorted, unrealistic beauty ideals.
Social media can further fuel body image concerns through negative feedback to user's own content and the images posted by others. In extreme cases, the negative feedback can manifest as body shaming and cyberbullying. Our study differs from existing literature in that it looks at social dynamics of online interactions that trap vulnerable individuals within toxic echo chambers.

\subsection{Online Pro-ana Communities}
Pro-ana communities are online spaces, originally on blogs, online forums, but also increasingly on social media platforms, that promote anorexia as a lifestyle not an illness. Pro-ana communities provide a venue for individuals with eating disorders to share tips on losing weight and concealing weight loss from others, as well as ``thinspriration'' images of very thin bodies to motivate weight loss~\cite{Ging2018}.
Researchers have argued that pro-ana communities have both positive and negative effects on individuals with eating disorders.
On the positive side, content analysis and qualitative studies revealed that these communities provide social support~\cite{juarascio2010pro} and a sense of belonging to individuals who often feel stigmatized and misunderstood~\cite{oksanen2016proanorexia,YeshuaKatz2013stigma}. Members can find empathy, encouragement, a safe space to vent, and information to help them better understand and manage their illness~\cite{McCormack2010}.
On the negative side, pro-ana communities often promote unhealthy and harmful behaviors that can exacerbate eating disorders, such as extreme calorie restriction, purging, and over-exercising. Members may compete with each other in weight loss, ask the group to hold them accountable to their weight loss goals or find ana buddies to go through the difficult periods of food restriction. The normalization of ED behaviors along with content that celebrates the ``thin ideal'' can increase psychological distress around body image and encourage disordered behaviors to persist~\cite{Mento2021}. Our paper extends these findings to social media platforms and links ED to other forms of deviant behaviors using online radicalization framework.

Computer scientists have studied online pro-anorexia communities.
\cite{pater2016hunger} characterized hashtags and media content associated with eating disorders-related posts on Tumblr, Instagram, and Twitter.  Distinct from this work, we also focus on how well integrated the eating disorders-related content is within the more mainstream diet and fitness content.
\cite{chancellor2016post} build a classifier to predict which posts will be taken down for violating Instagram's  prohibitions against promoting self-harm. The authors then combined  statistical text analysis of Instagram posts with clinician annotations to predict the severity of an individual
user's eating disorder~\cite{chancellor2016quantifying}.
In contrast to these works, we explore the social dynamics of pro-ana communities, including the mechanisms by which individuals  discover and become trapped within these communities. Rather than focus on individuals, our aim is to identify pathways for altering social dynamics so as to disrupt the growth of pro-ana communities.

\subsection{Online Radicalization}
Radicalization is a process by which individuals adopt extreme beliefs that enable them to engage in deviant behaviors, such as violence and terrorism. The internet has facilitated this process in a number of ways, creating an explosion of research into online radicalization. Social media platforms and discussion forums provide individuals access to online communities where they can receive social support and validation, engage with others who hold similar beliefs, and access extreme content that reinforces their beliefs and justifies extreme behaviors. In addition, algorithms on social media platforms may promote radicalization through the reinforcement of extremist beliefs through echo chambers and algorithmic amplification.

A number of theories have been proposed to explain the process of radicalization, focusing on political radicalization. These theories typically consider motivational, ideological, and social factors. \citet{Kruglanski2014} found that individuals who are more susceptible to online radicalization tend to be socially isolated and lack a sense of purpose or belonging~\cite{belanger2019radicalization}.

The \textit{3N model of radicalization} explains how violent extremists and terrorists are created~\cite{Kruglanski2022}. The model has three key elements: (\textit{i}) \textbf{Need for significance}, i.e., to feel that one matters or merits respect. Extremist ideology fills the void created by the loss of significance due to  humiliation, stigma or social identify that is disrespected by others;
(\textit{ii}) \textbf{Narratives} are justifications for using extreme behaviors (e.g., violence, terrorism) to pursue the goal for personal significance;
(\textit{iii}) \textbf{Networks} arise from social processes that connect individuals to like-minded others who validate their extremist ideology and narratives while isolating them from opposing ideas.
The 3N model has been empirically validated in different cultural settings in individuals expressing high levels of social isolation (need) and support for political violence (narrative)~\cite{belanger2019radicalization}. In addition, the 3N model was recently applied to explain emergence of online conspiracies like QAnon~\cite{wang2022identifying}.

The 3N model of radicalization provides a framework for understanding the growth of online eating disorders and self-harm communities. Individuals experiencing body image concerns may feel devalued and rejected, and are therefore vulnerable to being radicalized. Loss of personal significance motivates them to restore it~\cite{dugas2016quest} by  seeking out 
meaning and identity from extremist communities. These communities also provide narratives that promote and  justify extreme behaviors like food restriction, purging, or self-harm that restore personal significance, i.e., the thin ideal. Our work identifies elements of online activity linked to the 3N model and develops quantitative indicators to measure their effects.

\section{Methods}
\subsection{Data}
\label{sec:data}
Despite Twitter's policy prohibiting content that glorifies suicide and self-harm, tweets related to these topics have proliferated on the platform. One study found that the number of tweets mentioning  self-harm increased 500\% between 2021-10 and 2022-08~\cite{Goldenberg2022ncri}, with 20,000 monthly tweets on average. Twitter does not appear to enforce their content policy and rarely moderates harmful content. Although Twitter is adding links to mental health resources to search results for self-harm keywords, as of March 2023, searches for keywords related to eating disorders were not flagged.

We collected 2.532M tweets using keywords to query Twitter for messages covering the period from 2022-10 to 2023-03.
Our starting point was a set of terms identified by prior works~\cite{chancellor2016post,pater2016hunger} as promoting eating disorders. This include  \textit{thinspo} (``thininspiration''), \textit{proana} (pro-anorexia), and \textit{pro-mia} (pro-bulimia), among others. After examining results, we removed spammy terms like \textit{skinny} and added terms related to diet and weightloss topics such as (\textit{ketodiet}, \textit{weightloss}, $\ldots$), and anti-diet culture  (\textit{bodypositivity}, \textit{dietculture}, $\ldots$). See Appendix for the full list of keywords. Information collected on tweets includes user profile description, timestamp, message content, and  hashtags.


\subsection{Hashtag Co-occurrence Graph}
A hashtag co-occurrence graph is a network that represents relationships between hashtags in a  set of social media posts. An edge between two hashtags exists if they appear together in a post, with the weight of the edge representing the number of times--i.e., the number of tweets in which---the two hashtags co-occur. 

\subsection{Measuring Emotions}
Language carries cues to emotions, or feelings. These can include positive emotions, like joy and love, as well as negative emotions like anger and disgust. We use a state-of-the-art emotion detection model  SpanEmo~\cite{alhuzali2021spanemo}, which has been trained on SemEval 2018 Task 1 E-c data~\cite{SemEval2018Task1}. The SemEval dataset contains tweets labeled according to whether or not they contain emotions such as anger, disgust, fear, sadness, joy, love, optimism, pessimism and trust.  SpanEmo uses a transformer-based language model  to encode input text and returns a continuous value per emotion, denoting its confidence  that  the emotion is present in text.  Consequently, the model can predict none, one, or multiple emotions per tweet.

\begin{figure*}[tbh]
    \centering
    \includegraphics[width=0.85\linewidth]{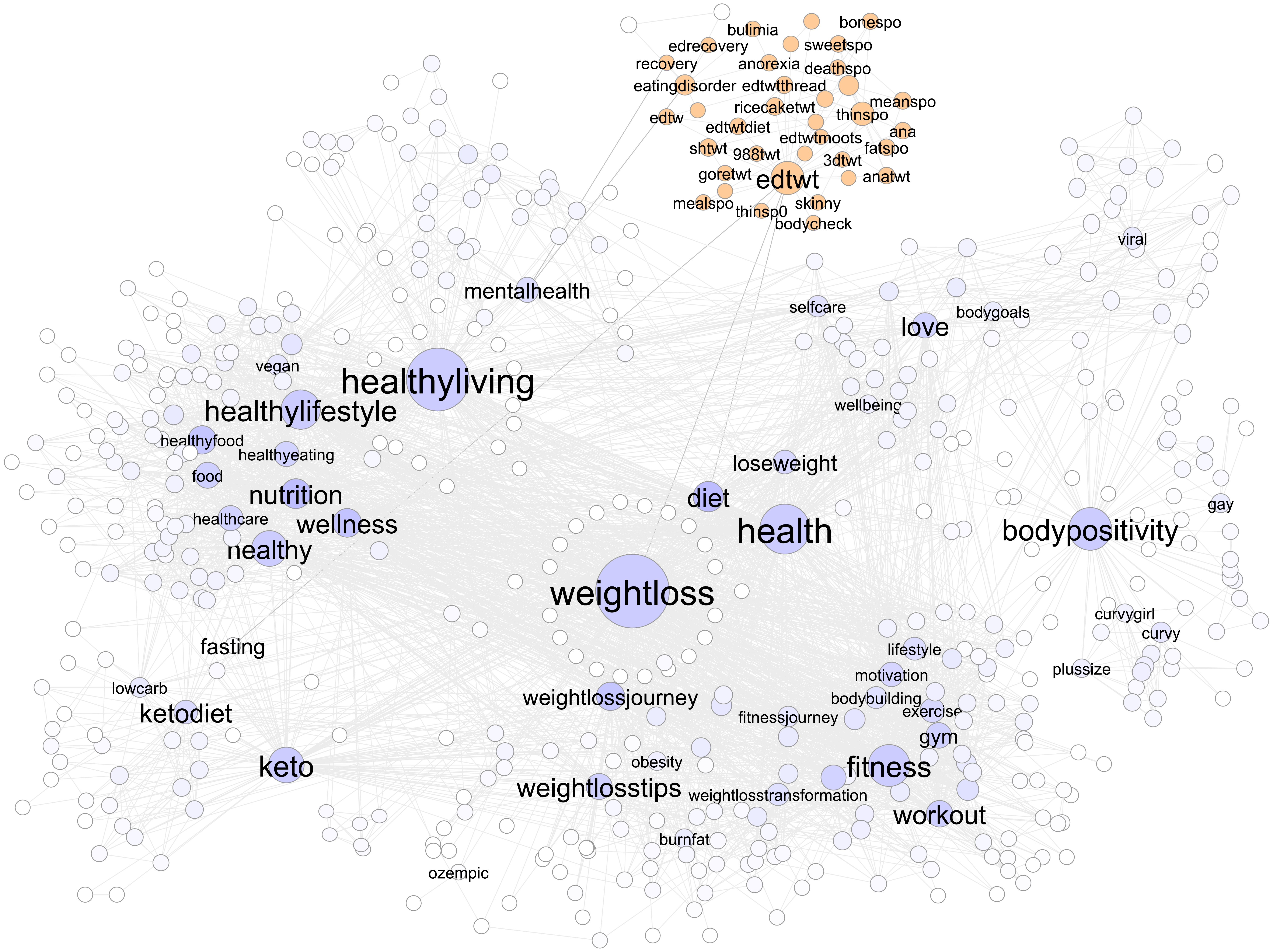}
    \caption{Hashtag co-occurrence network. Nodes represent the most popular hashtags (seen at least 230 times) in our data, with edges linking hashtags that frequently appear together in the same tweet (more than 100 times). Node size represents its centrality in the network. Harmful hashtags are highlighted light orange. }
    \label{fig:network}
\end{figure*}

\section{Results}
We use the methods described above to analyze social dynamics of online conversations about eating disorders.

\subsection{Accessibility of Pro-ana Communities}

One hazard of social media is that vulnerable individuals searching for health information may stumble upon harmful content promoting eating disorders. To investigate how well pro-ED discussions are integrated within mainstream health topics, we create a hashtag co-occurrence network of popular hashtags in our dataset. The hashtags represent important topics that users choose to highlight with a ``\#'' symbol (for clarity, we omit this symbol).  We focus on the 612 most popular hashtags in our dataset, which appeared more than 230 times.  Figure~\ref{fig:network} shows the largest connected component of the hashtag co-occurrence network  with 521 hashtags. Edges connect \textit{strongly linked} hashtags that co-occur in at least 100 tweets. Here node size represents its PageRank centrality and we used Louvain modularity to help visualize dense hashtag clusters. These clusters represent (1) \textit{Weightloss and fitness} community with topics such as diabetes, ozempic, workout, gym, fitspo, $\ldots$; (2) \textit{Keto diet} community with topics keto, paleo, lowcarb, $\ldots$;  (3) \textit{Healthy living} community with topics nutrition, wellness, vegan, cleaneating, mentalhealth, $\ldots$;
(4) \textit{Body positivity} community with topics curvy, thicc, plussize, $\ldots$;  and
(5) the pro-ED community highlighted in light orange. This cluster contains hashtags like \textit{thinspo}, a term used to describe content that glorifies extreme thinness,
\textit{proana},  a term used to refer to a subculture that views anorexia as a lifestyle choice rather than a serious illness,  and \textit{edtwt} and \textit{shtwt}, hashtags used to identify the ED and self-harm subcultures on Twitter, respectively.


ED-related hashtags are densely interconnected but they are also strongly linked  to four ``mainstream'' topics:  \textit{diet}, \textit{weightloss},  \textit{fasting}, and \textit{mentalhealth}.  When we relax the definition of ``strongly linked'' to mean hashtags that occur together in  30 (instead of 100) tweets, the ED hashtags remain well-segregated from other topics, forming a ``bubble.'' 
A concern here is that people could stumble upon ED content through popular gateway topics such as \textit{weightloss} and \textit{diet} and then become trapped in the bubble, isolated from the more mainstream views on nutrition, health and fitness.

We also find that eating disorder hashtags are strongly linked to other mental health conditions, especially  self-harm  (\textit{shtwt}), but also suicidal ideation (\textit{988twt}) and borderline personality disorder (\textit{bpdtwt}).
Interestingly, the self-harm topics clustered around the hashtag \textit{shtwt} (not shown) employ an array of euphemisms for cutting skin, the scars  it leaves, skin layers, and blood. These euphemisms are likely an effort to evade Twitter safety and moderation protocols, or perhaps an attempt to create an insider language. However, most of the connections between the two clusters are mediated by \textit{edtwt} and \textit{shtwt}, hinting at the central role they play in defining Twitter communities and allowing individuals to easily navigate between content created by these communities.



\subsection{Radicalization in Pro-ED Communities}
In this section, we operationalize the constructs of the 3N model of radicalization and link them to the patterns found in Twitter data.

\subsubsection{Need for Significance}

Loss of personal significance is a precipitating factor in radicalization~\cite{Kruglanski2014}. Studies of terrorism and violent extremism show that this can be triggered by a range of factors, such as marginalization, humiliation, social dislocation, identity crisis, despair, and political grievances~\cite{dugas2016quest,Kruglanski2022,belanger2019radicalization}. These factors motivate individuals to turn to extremism as a means of regaining significance. 
In the context of eating disorders, loss of significance may be triggered by negative body image and despair that drive vulnerable youth to seek dieting and weightloss advice online~\cite{Ging2018,choukas2022perfectstorm,pater2016hunger}.

Online eating disorder spaces fulfill the need for significance in several ways. They allow individuals to set \textit{body goals} (e.g., ``lose 20 lbs before prom'') and post progress pictures for \textit{accountability}.
They are also rife with ``inspirational'' images of grotesquely thin bodies that community members believe will motivate them and others to lose weight  or achieve a certain body shape. In fact, many hashtags in Fig.~\ref{fig:network} are related to ``thinspiration'', including \textit{thinspo}, \textit{thinsp0}, \textit{sweetspo}, \textit{bonespo}, \textit{deathspo}.

Another way that pro-ED communities help individuals regain significance is by provisioning them with a  collective identity~\cite{polletta2001collective}. Studies have shown that users self-label themselves in their online profiles, using popular hashtags to express solidarity with a chosen  group~\cite{barron2022quantifying} or cause, such as  ``\#blacklivesmatter'' or ``\#metoo''~\cite{chang2022justiceforgeorgefloyd}.
To check whether individuals adopt a collective identity of a pro-ED subculture, we check for the presence of strings ``edtwt'' and ``shtwt'' in their profile descriptions.
Of the 83,883 unique accounts in our dataset that posted 1,851,419 messages with harmful terms (\textit{edtwt}, \textit{shtwt}, \textit{thinspo}, etc.), 11,126 have ``edtwt'' in their profile description, and 3,217 have ``shtwt'' in  the description. Even without accounting for variations of these terms (e.g., ``ed twt''), 17\% of accounts tweeting about harmful topics were self-labeling themselves as belonging to the pro-ED and self-harm communities. This suggests that group identify is an important factor in joining harmful communities.

\subsubsection{Pro-ED Communities Promote Harmful Narratives}
Pro-ED tweets encourage and support harmful behaviors, such as restricting food intake, purging, and over-exercising, as a means for individuals achieve the ``thin ideal'' goal.
Among the set of 230 most popular hashtags, we identified those related to harmful topics that promote or enable harmful behaviors (excluding terms used for data collection). Table~\ref{tab:hashtags} categorizes these harmful hashtags.

\begin{table}
\caption{Harmful hashtags used in pro-eating disorders tweets categorized by topic.}
\label{tab:hashtags}
\hrule
\begin{description}
    \item[\textit{Motivation \& Inspiration}:]  \textit{Hashtags aiming to inspire others to engage in pro-ED behaviors with images that promote the ``thin ideal,'' or motivational images of progress (``bodycheck'') towards their ultimate goal weight (``ugw'').}
    \begin{enumerate}
        \item     thinsp0 thinspothread bonesp0 b0nesp0 b0nesp0 thinpo ribspo thinspiration thinspodaily th1nsp0 th1nspo
        \item     
    ugw thighgap  bodycheck bodychecking thinarms thinwaist ribs hipbones legs goals motivation
    \end{enumerate}

    \item[\textit{Information \& Tips}:] \textit{Hashtags used to label threads that provide detailed information  on how to  tolerate hunger,  avoid medical complications from starvation, or hide symptoms from caregivers. A thread may, for example, offer ``{things that help you /motivate you  to keep you }$\star${ving}'', details about ``{How I lost 7kg in a month},'' or recipes for an ``anorexia diet.''}
    \begin{enumerate}
        \item     edthread edthreads edtwtthreads edtips edtwttips restricting
        \item     
    anadiet eddiet edfood edtwtdiet edtwtfood edtwtfoodpoll edtwtrecipe abcdiet lowcal lowcalorie lowcalrecipe
    \end{enumerate}
    \item[\textit{Social \& Buddies}:] \textit{Hashtags used to request ``mutual'' followers or recruit ``buddies'' and  ``coaches'' to guide them through the anorexia journey.}
    \begin{enumerate}
        \item edmoots edmoot mutuals moots
        \item anabuddies anabuddy anacoach proanacoach
    \end{enumerate}

\end{description}
\hrule
\end{table}


\begin{table}[]
    \caption{The reach of pro-ED content. Engagement  with tweets (retweets and replies) on a variety of harmful topics promoting eating disorders and a control set of tweets.}
    \label{tab:groups}

    \centering
\small
    \begin{tabular}{c|c|c|c}
    \hline
    \textbf{Tweet group} & \textbf{Count} & \textbf{Retweets} & \textbf{Replies} \\ \hline
\textit{Motivation 1} & 70,749 & 101.12	&	0.35	\\
\textit{Motivation 2} & 157,610 & 132.44	&	0.53	\\
\textit{Information 1} & 5,799 & 114.80	&	0.57	\\
\textit{Information 2} & 23,573 & 97.88	&	0.32	\\
\textit{Social 1} & 128,381	& 43.39 &	0.70	\\
\textit{Social 2} & 1,803 & 4.68  &	1.07	\\
\textit{Random} & 10,000 & 139.91 & 0.52\\
\hline
    \end{tabular}
\end{table}

We grouped tweets according to the specified hashtags. We also created a random set of tweets from the data for comparison.
Table~\ref{tab:groups} reports the volume of tweets within each group and engagement levels, as measured by the average number of retweets and replies the tweets receive. Motivation and information tweets have wide reach, as demonstrated by large numbers of retweets. Requests for mutual followers are far more common than requests for pro-ana buddies and coaches, but the latter tweets receive more replies.
The level of engagement with pro-ED content is similar to that for the control set of tweets chosen at random from our dataset of tweets related to dieting and fitness. These results suggest that online communities play a role in  providing  information and inspiration to engage in disordered eating behaviors.

\subsubsection{Networks Offer Social Support}

Once individuals join an extremist group, a number of group dynamics processes keep them there~\cite{Kruglanski2014}. Similar processes also play  out within online pro-ED communities.  By linking individuals who are struggling with eating disorders to like-minded others, pro-ED  communities provide social and emotional support, as well as validation to those who may feel stigmatized and misunderstood. The desire to connect to like-minded others is visible in community introductions, where new members announce themselves while asking others to become their \textit{moots}, or ``mutual'' followers. The hashtag \textit{edtwtmoots} appears among the popular hashtags in Fig.~\ref{fig:network}.

We explore the means through which pro-ED communities provide support. First, we provide evidence that individuals identifying with pro-ED and self-harm subcultures form a tight community. Next, we show that members provide emotional support in the form of positive emotions.

\begin{table}[]
    \caption{Ratio of followers to followings for accounts that use specific terms in their profiles. We report the number of accounts using the specified term, the mean and standard deviation of the ratio of their followers/followings, a proxy ``mutual'' follows. Values near one suggest a high share of mutual followers, i.e., echo chambers. }
    \label{tab:ratio}
    \centering
    \small
    \begin{tabular}{c|c|c|c}
    \hline
\multicolumn{4}{c}{\textbf{User Followers/Followings Ratio}} \\ \hline
\textbf{Term} & \textbf{Count}  &   \textbf{Mean} &  \textbf{Std Dev} \\ 
\hline
\textit{edtwt} & 51,060  &  1.11  &  13.57 \\
\textit{shtwt} & 16,242  &  0.98  &   3.36 \\ 
\textit{diet}  &   3,217 &  51.74 & 1451.17 \\ 
\textit{weight} & 4,676  &  20.76 &  642.65 \\ 
\hline
    \end{tabular}
\end{table}

\paragraph{Social Embeddedness}
The group dynamics of online radicalization result in the formation of echo chambers that link individuals to like-minded others while isolating them from alternative points of view~\cite{barbera2020social}. 
Unfortunately, our dataset does not include information about the follower graph of users participating in eating disorders-related conversations, so we cannot formally test for the existence of echo chambers or the presence of reciprocated follower relationships, i.e., ``mutuals.'' However, our data includes user profile information, with user's bio, the number of followers as well as the number of accounts the user follows (followings). We use the ratio of  followers to followings as a proxy of the share of ``mutual'' followers. Table~\ref{tab:ratio} reports the mean value of this ratio for accounts that use one of the reported terms within their user profile. Accounts that use terms ``edtwt'' and ``shtwt'' as part of their profile have the ratio of followers to followings close to one, implying a large share of mutual connections. In contrast, accounts that use non-harmful terms ``diet'' and ``weight'' as part of their profile have a much larger ratio on average, albeit with very high variance. These results provide weak evidence that users identifying with ``edtwt'' and ``shtwt'' subcultures form echo chamber-like communities.

\begin{figure}
    \centering
    \includegraphics[width=0.8\linewidth]{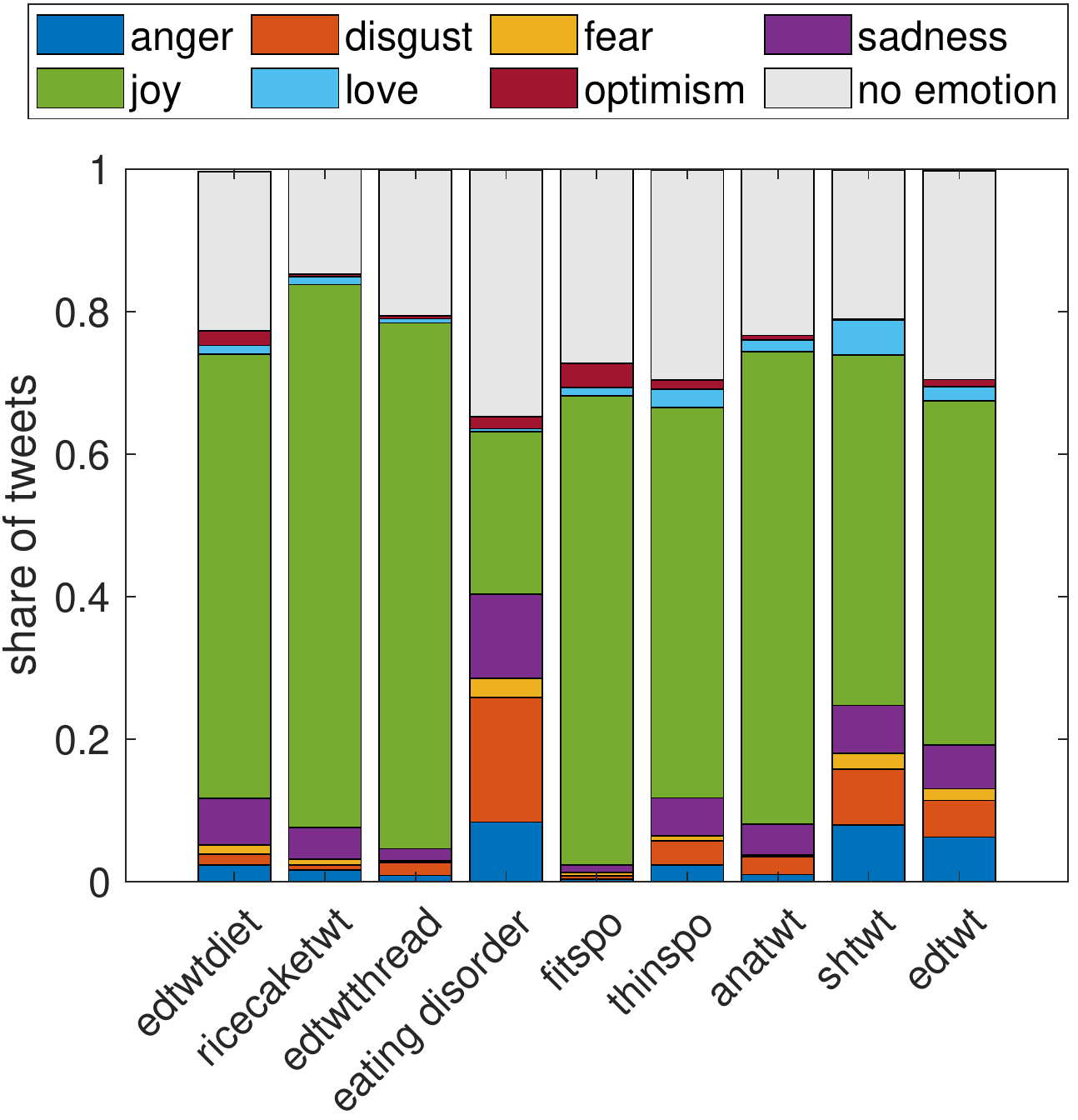}
    \caption{Emotion analysis. The share of tweets using specified terms that contain specific emotions.}
    \label{fig:emotions}
\end{figure}
\paragraph{Emotion Analysis}
We analyzed emotional expressions in tweets in our dataset. Joy was the most common emotion, occurring in almost 1.4M tweets, followed by optimism ($>$100K tweets), disgust (~100K), sadness, anger, fear and love. There was no emotion detected in a little more than 600K tweets.

Next, we focus on tweets with harmful terms. Figure~\ref{fig:emotions} shows the share of tweets mentioning select harmful terms that express each emotion. Joy is the most common emotion  in harmful tweets. 
Tweets mentioning `fitspo' (fit inspiration) and `edtwtthread', which includes tips that promote disordered eating behaviors, are predominantly joyful, as are community introductions where members request ``mutual followers.'' In contrast, tweets mentioning `eating disorder', `edtwt' and `shtwt'  express less joy and more negative emotions such as anger, disgust and sadness.  
The high prevalence of positive emotions within pro-ED messages suggests that communities  provide  emotional support. This may incentivize members to keep participating within ED communities.

\subsection{Harmful Tweet Classification}


Is it possible to automatically detect harmful tweets promoting eating disorders? To answer this question, we trained text classification models to recognize harmful tweets. We define three classification tasks: recognize harmful tweets that contain hashtags from 1)  the Motivation \& Inspiration category of Table~\ref{tab:hashtags}, 2) the Information \& Tips category, or 3) any harmful hashtag in Table~\ref{tab:hashtags}. These tasks allow us to  test how well we can recognize harmful tweets promoting thininspiration and over-excercising, as well sd harmful tweets that promote extreme dieting.

To represent the text of the tweets (excluding hashtags) we used bag-of-words (TF-IDF) representation and word embeddings (Word2Vec).   For consistency, we kept the number of features in the TF-IDF representation at 50K, and set the vector size in word2vec at 100 across all models. We used three classification models---logistic regression, SVM, and XGBoost---and used default parameters in training all  models. For each classification task, we created a balanced dataset of 50K randomly chosen harmful tweets and 50K random non-harm tweets.
We trained all models on the randomly selected 80\%  of tweets and tested classification performance on the held out 20\% of tweets.


\begin{table}[]
    \caption{Accuracy of harmful tweet classification.}
    \label{tab:predictions}

    \centering
    \small
    \begin{tabular}{l|c|c|c}
    \hline
	&	\textbf{motivation} 	&	\textbf{information}	&
 \textbf{all}\\ \hline
tfidf logreg	&	80.99	&	90.57	& 86.39\\
tfidf SVM	&	80.27	&	88.39	& 86.53\\
tfidf XGBoost	&	79.46	&	89.03	& 86.94\\
word2vec logreg	&	76.61	&	86.65	& 85.35\\
word2vec SVM	&	77.57	&	89.58	& 85.16\\
word2vec XGBoost	&	80.37	&	91.88	& 86.51\\
\hline
    \end{tabular}
\end{table}

Table~\ref{tab:predictions} reports the accuracy of harmful tweet classification on the three tasks. Across both harmful motivation and information tweets, logistic regression using TF-IDF representation and XGBoost using word embeddings achieved best performance, distinguishing harmful tweets from non-harmful ones with 90\% accuracty.
We then combined the two datasets to create a balanced dataset of 100K harmful tweets and 100K non-harmful tweet and report classification performance. In this case, XGBoost gives best performance of 87\% with both representations.
These results suggest that it is relatively easy for social media platforms to distinguish harmful tweets promoting eating disorders from non-harmful tweets about fitness and diet.

\section{Conclusion}
Our paper examines the role of social media platforms like Twitter in eating disorders. We argue that the growth of online communities that promote eating disorders and other forms of self-harm can be described as online radicalization process~\cite{Kruglanski2014,Kruglanski2022,belanger2019radicalization}. This helps explains how vulnerable individuals seek  meaning and social connection in communities that  adoption of extreme worldviews. Although the model describes offline behaviors that create terrorists and violent extremists, it was also applied to online conspiracies, hate groups, and extremist ideologies.
To the best of our knowledge, our work is the first to apply this model to mental health conditions like eating disorders. By facilitating social connections to like-minded others, a shared group identity and a sense of belonging, pro-anorexia communities create toxic echo chambers that normalize extreme behaviors that lead to eating disorders  and self-harm.
These communities  can make recovery difficult.


Our work highlights the importance of social dynamics in explaining mental health trends and contributes to the body of research implicating social media in the transmission of behaviors that lead to negative mental health outcomes. The ongoing mental health crisis among the youth creates an urgent need for solutions. The radicalization model explored in this paper provides a framework for understanding the phenomenon, links it to other harmful behaviors, and provides insights into potential mitigation strategies. At the very least, our work calls for more stringent moderation of online content that is harmful to mental health.






\begin{acks}

\end{acks}

\bibliographystyle{ACM-Reference-Format}
\bibliography{references}

\appendix

\section{Search Terms}

The keywords that we use to retrieve ED tweets are: \emph{thinspo}, \emph{edtwt}, \emph{proana}, \emph{proanatips}, \emph{anatips}, \emph{meanspo}, \emph{fearfood}, \emph{sweetspo}, \emph{eatingdisorder}, \emph{bonespo}, \emph{promia}, \emph{redbracetpro}, \emph{bonespo}, \emph{m34nspo}, \emph{fatspo}, \emph{lowcalrestriction}, \emph{edvent}, \emph{WhatIEatInADay}, \emph{Iwillbeskinny }, \emph{thinspoa }, \emph{skinnycheck }, \emph{thighgapworkout}, \emph{bodyimage}, \emph{bodygoals}, \emph{weightloss}, \emph{skinnydiet}, \emph{chloetingchallange}, \emph{fatacceptance}, \emph{midriff}, \emph{foodistheenemy}, \emph{cleanvegan}, \emph{keto}, \emph{ketodiet },  \emph{cleaneating}, \emph{intermittentfasting}, \emph{juicecleanse}, \emph{watercleanse}, \emph{EDrecovery}, \emph{bodypositivity}, \emph{dietculture}.


\end{document}